\documentclass[12pt,twoside,fleqn]{article}

\usepackage{epsfig}
\usepackage{psfig}
\def\lsim{\raise0.3ex\hbox{$<$\kern-0.75em\raise-1.1ex\hbox{$\sim$}}}
\def\gsim{\raise0.3ex\hbox{$>$\kern-0.75em\raise-1.1ex\hbox{$\sim$}}}
\makeatletter
\@addtoreset{equation}{section}
\makeatother

\newcommand{\beqn} {\begin{equation}}
\newcommand{\eqn} {\end{equation}}
\newcommand{\eqa}{\begin{eqnarray}}
\newcommand{\ena}{\end{eqnarray}}

\newcommand{\slsh}[1] {#1\kern-.43em/}

\newcommand{\real}{{\sf I}\kern-.12em{\sf R}}
\newcommand{\comp}{{\sf I}\kern-.48em{\sf C}}

\newcommand{\aleq}{\mbox{}_{\textstyle \sim}^{\textstyle < }}

\newcommand{\nin} {\in\kern-.6em/}

\def\MEF{m_{\rm eff}}\def\mef{\ifmmode\MEF\else$\MEF$\fi}

\def\NT{N_\tau}
\def\nt{\ifmmode\NT\else$\NT$\fi}
\def\NS{N_\sigma}
\def\ns{\ifmmode\NS\else$\NS$\fi}

\def\PR{{ Phys.\ Rev.\ }}

\def\PL{{ Phys.\ Lett.\ }}
\def\NP{{ Nucl.\ Phys.\ }}
\def\OP{{\langle M \rangle }}
\def\OPQ{{\langle M^2 \rangle }}
\def\MO{{\langle |M| \rangle }}

\begin{document}
\thispagestyle{empty}
%
 \mbox{} \hfill BI-TP 98/23~\\
 \mbox{} \hfill August 1998\\
\begin{center}
\vspace*{1.0cm}
{{\large \bf The Calculation of Critical Amplitudes \\
         in SU(2) Lattice Gauge Theory}
 } \\
\vspace*{1.0cm}
{\large J. Engels and T. Scheideler } \\
\vspace*{1.0cm}
{\normalsize
$\mbox{}$ {Fakult\"at f\"ur Physik, Universit\"at Bielefeld,
D-33615 Bielefeld, Germany}}\\
\vspace*{2cm}
{\large \bf Abstract}
\end{center}
\setlength{\baselineskip}{1.3\baselineskip}

We calculate the critical amplitudes of the Polyakov loop and its
susceptibility at the deconfinement transition of (3+1) dimensional
$SU(2)$ gauge theory.
To this end we study the corrections due to irrelevant exponents
in the scaling functions. As a guiding line
for determining the critical amplitudes we use envelope equations 
which we derive from the finite size scaling formulae of the observables. 
We have produced new high precision data on $\ns^3\times4$
lattices for $\ns=12,18,26$ and 36. With these data
we find different corrections to the asymptotic scaling behaviour
above and below the transition. Our result for the 
universal ratio of the susceptibility amplitudes
is $C_+/C_-=4.72(11)$ and thus in
excellent agreement with a recent measurement for the $3d$ Ising model.

\newpage

\setcounter{page}{1}
\vspace*{1cm}

\section{Introduction}


The calculation of critical exponents at the critical point of second order 
transitions with
Monte Carlo methods is by now standard. To this end one has to simulate
the theory under consideration in
different volumes in the immediate neighbourhood of the transition point.
The behaviour of thermodynamic quantities in the thermodynamic limit
is then inferred from extrapolation formulae which are 
derived from finite size scaling (FSS) theory. This allows in principle a 
classification of the underlying theory,
because the critical exponents are universal for all models belonging to
the same universality class. Yet the differences between the 
exponents of different classes may be rather small. 
Further tests on universality should then be performed. Indeed, since 
members of the same class are also sharing various
scaling functions it can be shown that certain critical point 
amplitude combinations are universal as well \cite{AHP91}. 
Their calculation using 
finite volume simulations is, however, far more demanding,
than in the case of the critical indices. The reason for this is that
the amplitudes have to be taken from the $tL^{1/\nu}\rightarrow\infty$ 
limit of the scaling functions, where $t~$is the reduced temperature and 
$L$ is the characteristic length scale, $L=V^{1/d}$, 
i.e. essentially from very large volumes.  

Recently Caselle and Hasenbusch \cite{Case97} were able to show that it
is possible to obtain Monte Carlo estimates for critical point amplitude
ratios in the $3d$ Ising model with a precision comparable to those of 
other approaches \cite{BLZ74}$-$\cite{Bute98}. 
In the more complex $SU(2)$ gauge theory in 
(3+1) dimensions, which is a member of the same universality class, this 
has been a dream for quite some time. An early attempt \cite{Joe92} to deduce
information on amplitudes from Monte Carlo data in $SU(2)$ led to the 
conclusion, that the existing data were still inadequate for meaningful 
comparisons to results from analytic calculations. As we shall see later, both 
the quality of the data and the method of determination of the amplitudes 
are of great importance for the success of the project. There are other
difficulties : the critical point has to be known with high accuracy, because
a shift changes the estimates of the scaling functions. In the
$3d$ Ising model the critical point has been determined with extreme
precision, and simulations on really large lattices - up to $L=120$ in 
\cite{Case97} - have been performed. Such lattice sizes  are still out of
reach for $SU(2)$ calculations.

 In view of this situation we have chosen a different method from Caselle and
Hasenbusch. We proceed in the following way. In section 2 we describe
how one can control the approach to the asymptotic scaling form. For this
purpose we consider the envelope function to the family of curves, which
one obtains for different volumes.
In the following section we present our data. Section 4 contains the analysis.
Here, we first ascertain again the location of the critical point \cite{Eng96}
with the new data, then we study the scaled observables and examine the 
corrections to the scaling functions. The critical amplitudes are finally
derived from the estimates of the corrected scaling functions. We close with 
a summary and the conclusions.


\section{The Approach to the Asymptotic Scaling Form}

\subsection{Critical point amplitudes}


To stay as general as possible we use in this section the notation for
magnetic systems. We define the reduced temperature $t$ as
\beqn
t={T-T_c \over T_c},
\eqn
where $T$ ist the temperature and $T_c$ the critical temperature. 
In the thermodynamic limit the correlation length 
$\xi$ diverges at a second order transition as
\beqn
\xi=f_{\pm}|t|^{-\nu}.
\eqn
Here the index of the critical amplitude $f$ refers to the symmetric (+)
or to the broken phase ($-$) and coincides for magnetic systems with the sign 
of $t$. The magnetization or order parameter $\OP$ and the magnetic 
susceptibility $\chi$ behave for zero external magnetic field $H$ close to the 
critical point as follows
\beqn
\OP=B(-t)^{\beta}~~ {\rm for~} t<0~,
\eqn
and
\beqn
\chi=C_{\pm}|t|^{-\gamma}.
\eqn
Though the amplitudes $C_+$ and $C_-$ are not universal, their ratio is.
The same is true for $f_+$ and $f_-$. More universal amplitude ratios are
obtained by making use of the hyperscaling relations among the critical 
exponents. 

\subsection{Finite size scaling}


The approach to the just mentioned asymptotic scaling forms of the 
thermodynamic quantities is described by finite size scaling equations.
In particular, it can be shown \cite{Barb83} using renormalization 
group theory that the singular part of the free energy density has the 
form
\beqn
f_s(t,H,L) = L^{-d}Q_{f_s}(g_T L^{1/\nu},
g_H L^{{(\beta + \gamma )}/ \nu}, g_i L^{\lambda_i})~.
\eqn
The scaling function $Q_{f_s}$ depends on the temperature $T$ and the
external field strength $H$ in terms of a thermal and a magnetic
scaling field
\beqn
g_T = c_T t + O(tH,t^2)~,
\label{gT}
\eqn
\beqn
g_H = c_H H + O(tH,H^2)~,
\eqn
and possibly further irrelevant scaling fields $g_i$ with negative exponents
$\lambda_i$. All scaling fields are independent of $L$.

The order parameter $\OP$, the susceptibility $\chi$ and
the normalized fourth cumulant $g_r$ of the magnetization
\beqn 
g_r = { {\langle M^4 \rangle} \over {\OPQ}^2 } - 3~,
\eqn
are obtained from $f_s$ by taking
derivatives with respect to $H$ at $H=0$.
 The general form of the scaling relations derived in this way 
for an observable $O$ is
\beqn
O(t,L) = L^{\rho / \nu} \cdot {\bar Q}
(g_T L^{1/{\nu}},g_i L^{\lambda_i})~.
\eqn
Here $O$ is $\OP, \chi$ or $g_r$ with
$\rho = - \beta, \gamma~{\rm and}~ 0$, respectively.
Taking into account only the largest irrelevant exponent $\lambda_1=-\omega$
and inserting the expansion \ref{gT} into $\bar Q$ we arrive for small $|t|$ 
at
\beqn
O(t,L) = L^{\rho / \nu} \cdot Q(tL^{1/{\nu}},L^{-\omega})~.
\label{FSS}
\eqn


\subsection{Control of approach to the thermodynamic limit}

\begin{figure}[htb]
\begin{center}
   \epsfig{bbllx=85,bblly=250,bburx=495,bbury=560,
       file=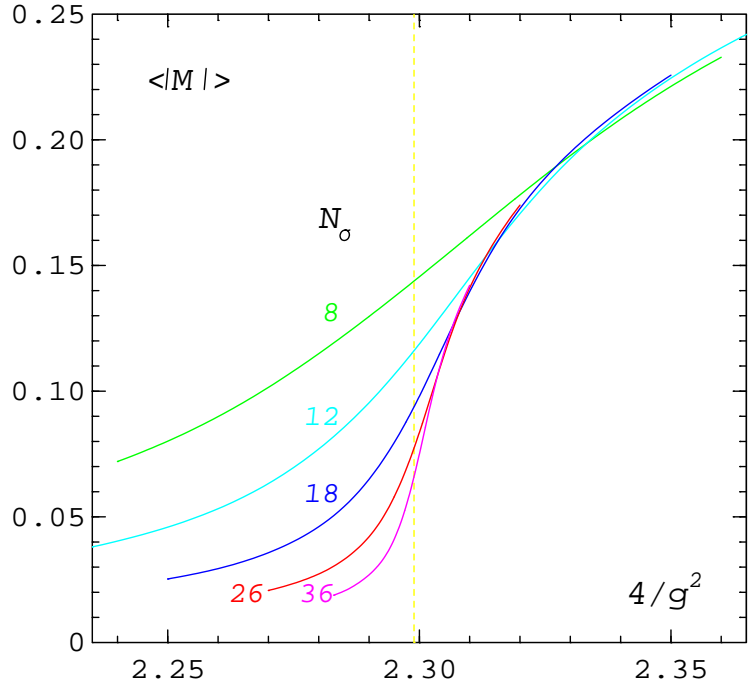, width=110mm,height=80mm}
\end{center}
\caption{The modulus of the Polyakov loop $\MO$ as function of $4/g^2$
on $\ns^3\times 4$ lattices. The dashed line shows the location of the
critical point.} 

\label{fig:labs}
\end{figure}

 The functions $O(t,L)$ for a specific observable build a family of curves, 
parametrized by $L$. For increasing $L$ these functions are supposed to
approach the limiting form
\beqn
O_{\infty}(t) = a_0 |t|^{-\rho} {\rm~~for~~} |t|\rightarrow 0.
\label{limit}
\eqn 
An inspection of such an ensemble of curves from Monte Carlo measurements
on different volumes suggests that one calculate the envelope
function to the family of curves. An example of this is
the magnetization in $SU(2)$ shown in Fig.~\ref{fig:labs}. 
\vskip 0.8truecm
\newpage
At least the leading term in $t$ of 
this function should coincide with the limiting form eq. \ref{limit}.
The amplitude could then be determined from the envelope function.  

We derive the envelope function from the FSS formula \ref{FSS}
\beqn
F_e = O L^{-\rho/ \nu}-Q =0,
\label{first}
\eqn
by solving the equation
\beqn
{\partial F_e \over \partial L} = 0
\label{second}
\eqn
for $L=L(t)$ and insertion into eq. \ref{first}. Here $L(t)$ defines
the matching point $t$, where the envelope function touches the curve
with parameter $L(t)$. 
The scaling function $Q$ depends only on the scaled reduced temperature $x$ 
and the correction-to-scaling variable $y$
\beqn
x = tL^{1/\nu},~~y = L^{-\omega}~.
\eqn
Eq. \ref{second} can then be written as
\beqn
0 = \rho Q + x {\partial Q \over \partial x} - \omega \nu y {\partial Q 
\over \partial y}~.
\label{third}
\eqn
In the following we assume a linear dependence of $Q$ on $y$ 
\beqn
Q(x,y) = Q_0(x) + y Q_1(x)~,
\eqn
which is certainly justified for large $L$. 
We will check our data for this point. Inserting the
last equation into eq. \ref{third} leads to
\beqn
0=\rho Q_0(x)+x Q_0^{\prime}(x)+y \left[ (\rho-\omega\nu)Q_1(x) 
+xQ_1^{\prime}(x)\right]~.   
\eqn
 The last equation can be solved in a first approximation for $y=0$
by determining $x_0$ from
\beqn
0=\rho Q_0(x_0) + x_0 Q_0^{\prime}(x_0)~,
\label{solve}
\eqn 
which corresponds to the approximate matching point 
\beqn
L_0 = \left ( {x_0 \over t} \right)^{\nu}.
\eqn
The next approximation $L_1$ is obtained from the ansatz
\beqn
L_1^{1/\nu} = {x_0 \over t} (1+\epsilon(t)) = {x_1 \over t}~~,
\eqn
with the result
\beqn
\epsilon (t) = { (t/x_0)^{\omega\nu} \tilde Q_1 \over
\rho(\rho+1)Q_0 -x_0^2 Q_0^{\prime\prime} -(t/x_0)^{\omega\nu}
(\omega\nu \tilde Q_1+x_0 \tilde Q_1^{\prime})}~~,
\eqn
where
\beqn
\tilde Q_1 = (\rho-\omega\nu)Q_1+x_0Q_1^{\prime}~~,
\eqn
and all $Q$ have to be taken at $x=x_0$.

Inserting $L_1(t)$ into eq. \ref{first} gives the envelope function 
\beqn
O_e = ({t \over x_0})^{-\rho}\left\{ Q_0(x_0) + ({t \over x_0})^{\omega\nu}
Q_1(x_0) + O(|t|^{2\omega\nu}) \right\}~.
\label{enve}
\eqn
The sign of $t$ and $x_0$ are here the same, the envelope function exists 
only on that side of the transition, where a solution $x_0$ to eq. \ref{solve}
is found. We 
note that the correction term $\epsilon(t)$ does not enter the first
correction-to-scaling term in $O_e$. The form of $O_e$ is the same as the one
expected for $O_{\infty}$, eq. \ref{limit}. Also the correction-to-scaling
term has the correct exponent, namely
\beqn
\theta = \omega\nu~.
\eqn
Comparing the expressions \ref{limit} and \ref{enve} we find
\beqn
a_0 = |x_0|^{\rho} Q_0(x_0)~.
\eqn 
Of course, this result does not come as a surprise. Suppose, there are no 
scaling corrections, so that
\beqn
O(t,L) = L^{\rho/\nu} Q_0(x) = |t|^{-\rho} |x|^{\rho} Q_0(x)~.
\eqn
As a consequence we get in the thermodynamic limit
\beqn
O_{\infty} = |t|^{-\rho} \lim \limits_{ x \to \infty} |x|^{\rho} Q_0(x)~.
\eqn
Consider now the function 
\beqn
A_0(x) = |x|^{\rho} Q_0(x)~,
\eqn
and its approach to asymptopia. Its derivative is given by
\beqn
A_0^{\prime}(x) = {\rm sign}(x)|x|^{\rho-1} (\rho Q_0+xQ_0^{\prime})~.
\eqn
The bracket expression in the last equation becomes zero at the 
matching point $x_0$ and also if $Q_0(x)\sim |x|^{-\rho}$, that is 
when it reaches its asymptotic form. Then $A_0(x)$ attains an
extreme value, the critical point amplitude $a_0$. 

 From the above considerations we deduce our method of calculation
of the critical point amplitudes. In a first step, we estimate
the scaling function $Q_0(x)$ from the data, by carefully examining
the corrections-to-scaling contributions to $Q(x,y)$. Next we control 
the approach to the correct scaling form of $Q_0(x)$, by calculating
the function
\beqn
F_O = \rho Q_0+x Q_0^{\prime}~.
\label{Control}
\eqn 
It should become zero inside the error bars if $x$ is large enough. A
single zero at small $x$ is obviously not what we are looking for.


\section{$SU(2)$ gauge theory in (3+1) dimensions}


In the following we consider $SU(2)$ gauge theory on $\ns^3\times\nt$
lattices, where $\ns$ and $\nt$ are the number of lattice points in the space
and time directions. Volume and characteristic length scale $L$ are given by 
\beqn
V = (N_\sigma a)^3~,~~~ L = \ns a~.
\eqn
Here, $a$ is the lattice spacing. 
For all practical purposes we can take $a=1$, so that $L$ and $\ns$ are
equivalent. We use the standard Wilson action
\beqn
S(U) =  {4\over g^2} \sum_p (1 -{1\over 2}Tr U_p)~,
\eqn
where $U_p$ is the product of link operators around a plaquette.
In contrast to magnetic systems, where the phase of spontaneous
magnetization or symmetry breaking is at physical temperatures $T<T_c$, 
the situation at the deconfinement transition is just reverse: the
symmetric phase is below $T_c$. Correspondingly the sign of the 
reduced temperature belonging to a certain phase is opposite to the usual
one. The reduced temperature may be approximated in $SU(2)$ near the 
transition through
\beqn
{\bar t} = { 4/g^2 - 4/g^2_c \over 4/g^2_c }~,
\eqn 
where $4/g^2$ is the coupling constant and we have denoted the reduced
temperature with $\bar t$ to keep the sign difference in mind. 

On an infinite volume lattice the order parameter or
magnetization for the
deconfinement transition is the expectation value of the Polyakov loop
\beqn
M({\bf x}) =  {1\over 2} Tr \prod_{\tau=1}^{N_\tau}U_{\tau,
{\bf x};4}~,
\eqn
or else, that of its lattice average
\beqn
M = {1\over V}\sum_{\bf x}M({\bf x})~,
\eqn
where the $U_{\tau,{\bf x};4}$ are the $SU(2)$ link matrices in time 
direction. Due to system flips between the two ordered states on
finite lattices the expectation value $\OP$ is always zero. Therefore
we replace it by the expectation value of the modulus of
the magnetization, $\MO$.
This observable was shown to converge to the correct infinite volume value
in the broken phase at least for the $3d$ Ising model \cite{Case97,Tala96}.
The FSS investigations in $SU(2)$, which used this observable (see e.g.  
\cite{Eng96}) confirmed this finding. Correspondingly we use instead of
the true susceptibility the definition
\beqn
\chi = V (\OPQ - \MO^2)~.
\eqn
In the symmetric phase, however, the finite volume susceptibility
\beqn
\chi_v = V \OPQ~,
\eqn
is the appropriate choice. At the critical point $\bar t =0$, the data for both 
$\chi$ and $\chi_v$ show FSS behaviour with the same critical exponent  
$\gamma$.


\subsection{The data}

\begin{figure}[htb]
\begin{center}
   \epsfig{bbllx=85,bblly=250,bburx=495,bbury=560,
       file=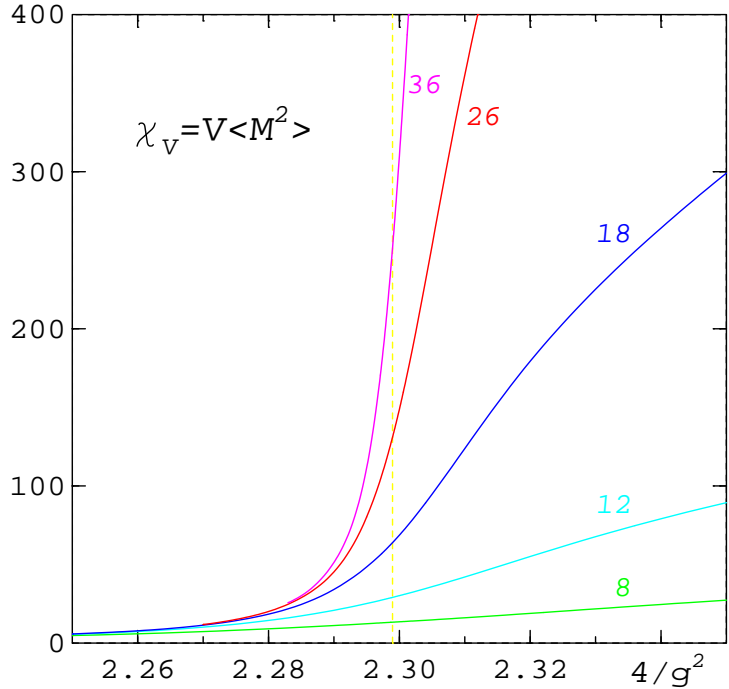, width=110mm,height=80mm}
\end{center}
\caption{The finite volume susceptibility $\chi_v$ as function of $4/g^2$.
The numbers indicate $\ns$ and the vertical 
line shows the location of the critical point.}

\label{fig:cv}
\end{figure}
\begin{figure}[htb]
\begin{center}
   \epsfig{bbllx=50,bblly=50,bburx=550,bbury=750,
       file=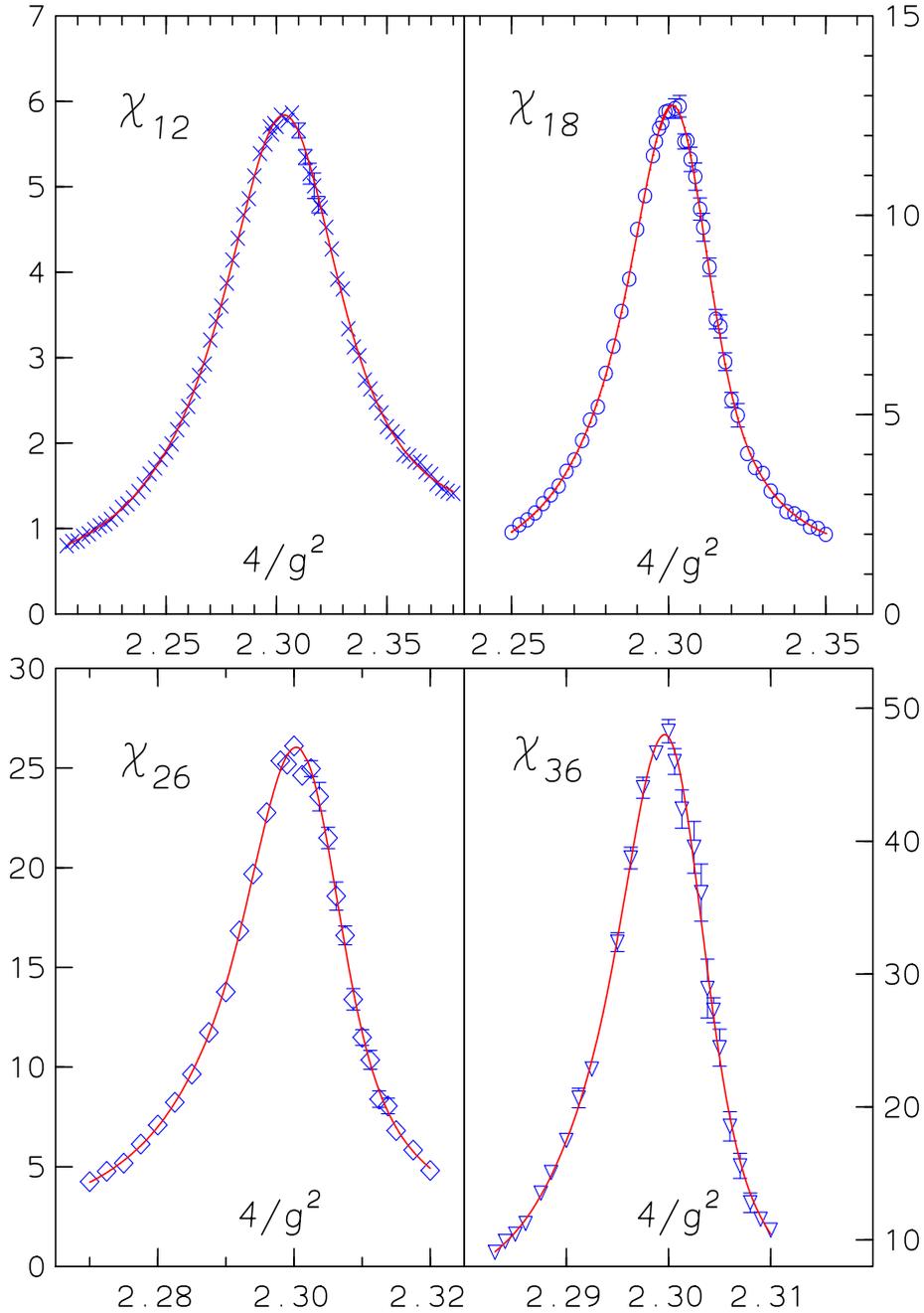, width=120mm,height=180mm}
\end{center}
\caption{The susceptibility $\chi$ as function of $4/g^2$.
The index on $\chi$ is $\ns$, the lines are the DSM interpolations.}

\label{fig:chidat}
\end{figure}

Originally we started our analysis with Monte Carlo data from $\ns^3 \times 4$
lattices, which we took from refs. \cite{Eng96} and \cite{Eng92}. They were
well suited for the determination of the critical indices and the critcal
coupling $4/g^2_c$. It turned however soon out that they were not precise 
enough to reliably estimate the scaling function $Q_0(x)$ and secondly that 
we needed data in a larger range of $x-$values. We have therefore produced 
four complete new sets of data on $\ns^3 \times 4$ lattices with $\ns=12,18,26$
and 36 on our workstation cluster. Between the measurements five updates, 
consisting of one heatbath and two overrelaxation steps were performed. 
Compared to the old data the integrated autocorrelation time $\tau_{int}$ 
is now considerably reduced. The minimal number of measurements 
per coupling was 20000,
close to the critical point between 40000 and 80000. The different 
coupling values were so densely chosen, that their plaquette distributions
were overlapping to a large extent. It was therefore easy to apply the
density of states method (DSM)\cite{DSM} in the whole range. Our subsequent
analysis of the data will be based on their DSM interpolations. 
A general survey of our data is given in 
Table~\ref{tab:survey}. In the appendix we list the results in detail.  
\begin{table}
\begin{center}
\begin{tabular}{|r|c|c|c|c|c|}
\hline
$\ns$ & $4/g^2{\rm -range}$ & ${\rm No.}(4/g^2)$  
& \multicolumn{1}{|c|}{$\tau_{int}(\bar t<0)$ }
& \multicolumn{1}{|c|}{$\tau_{int}(\bar t\approx 0)$ }
& \multicolumn{1}{|c|}{$\tau_{int}(\bar t>0)$} \\
\hline
   12    & 2.205-2.38 &   74    &   3-5        &   ~8-12      &  4-7        \\
   18    & 2.25 -2.35 &   49    &   3-9        &   10-20      &  3-9        \\
   26    & 2.27 -2.32 &   29    &   3-9        &   15-40      &  ~7-20    \\
   36    & 2.283-2.31 &   26    &   3-10       &   20-70      &  15-40    \\
\hline
\end{tabular}
\end{center}
\caption{Survey of the Monte Carlo simulations for different lattices. 
Here No.$(4/g^2)$ is the number of different couplings at which
runs were performed; $\tau_{int}$
is the integrated autocorrelation time for the measured
plaquette lattice averages. }
\label{tab:survey}
\end{table}
In Fig. \ref{fig:labs} we showed already the DSM interpolations to our 
data for the 
modulus of the Polyakov loop, in Fig. \ref{fig:cv} 
the corresponding ones are plotted for $\chi_v$ 
All figures contain also previous results for $\ns=8$ from \cite{Eng92}.
In order to give an impression of
the amount and quality of our new data, we present in Fig. \ref{fig:chidat} the
results for the directly measured data points for the susceptibility $\chi$.
We note that the susceptibility has much larger statistical errors than 
$\MO$ and $\chi_v$. 

\vskip 1.0truecm
\newpage

\begin{figure}[htb]
\begin{center}
   \epsfig{bbllx=35,bblly=240,bburx=550,bbury=615,
       file=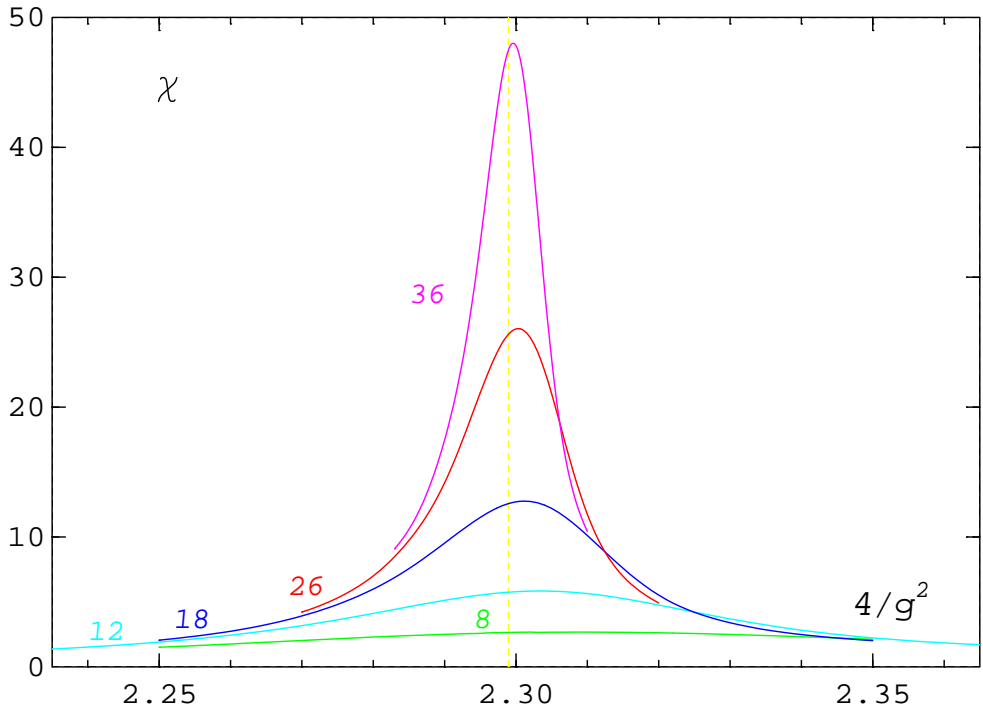, width=140mm,height=100mm}
\end{center}
\caption{The susceptibility $\chi$ as function of $4/g^2$.
The numbers are $\ns$, the lines are the DSM interpolations and 
the dashed line shows the location of the critical point.}

\label{fig:chi}
\end{figure}

As was the case for the $3d$ Ising model \cite{Case97}, the simulations in 
the symmetric phase required fewer measurements for the same accuracy. This 
is observed also in Fig. \ref{fig:chidat}, where the errors for $\bar t>0$ 
are larger than for $\bar t<0$, though we made in general more measurements 
there. In Fig. \ref{fig:chi} we compare the results for $\chi$ for the
different volumes.


\section{Scaling analysis of the data}
 

\subsection{The critical point}


For our scaling analysis it is important to know the exact location of the
critical point. We have therefore repeated the determination of the critical
point with our new data and the $\chi^2-$method as proposed in ref. 
\cite{Eng96}. That method is a test on the $L-$dependence of an observable 
$O$ at the critical point $t=0,x=0$
\beqn
O(t=0,L) = L^{\rho / \nu} \cdot \left(Q_0(0)+L^{-\omega}Q_1(0)\right)~.
\label{chi}
\eqn

\vskip 1.0truecm
\begin{figure}[htb]
\begin{center}
   \epsfig{bbllx=85,bblly=250,bburx=495,bbury=560,
       file=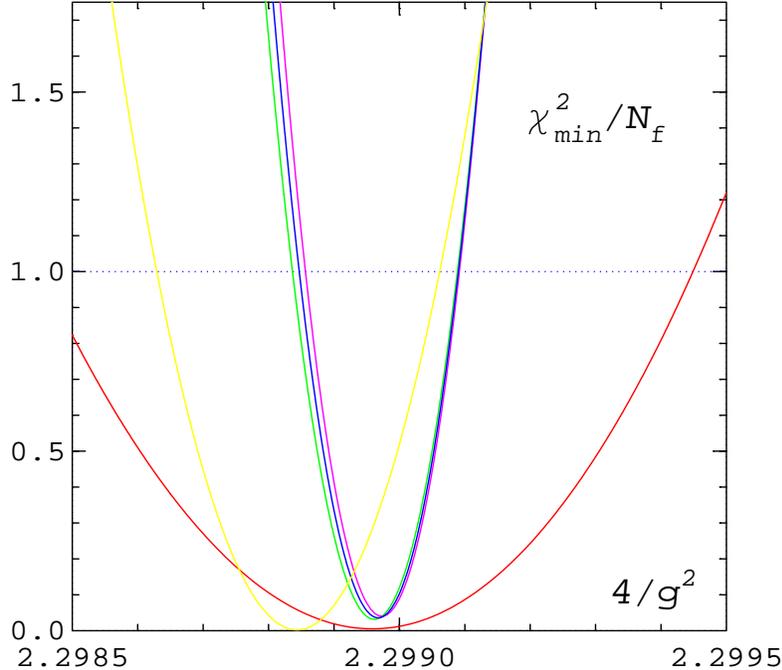, width=110mm,height=80mm}
\end{center}

\caption{The minimal $\chi^2$ per degree of freedom for fits of $\MO$ 
at fixed $4/g^2$ according to eq. \ref{chi}: from \cite{Eng96}(red);
fit of new data, leading term only(yellow) and full fit with $\omega=1.1,1.2,1.3$
(green,blue,magenta).  }

\label{fig:crip}
\end{figure}

If $x\not=0$ the $L-$dependence is drastically changed. At the critical point
a fit to the form \ref{chi} has therefore the least minimal $\chi^2$ .
Taking into account only the leading term in \ref{chi}, a fit of ln$O$ as a
function of ln$L$ gives at the same time the value of $\rho/\nu$. The results
for $\beta/\nu$ and $\gamma/\nu$ in \cite{Eng96} were each about 1$\%$ off 
the values calculated from the Ising exponents used in ref. \cite{Case97}
          \begin{equation}
          \beta = 0.327~,~~\gamma = 1.239~,~~\nu = 0.631~,
          \label{crex}
          \end{equation}
though the hyperscaling relation $d=2\beta/\nu +\gamma/\nu~$
was fulfilled up to 0.2$\%$. The set of data we have used in the current critical
point analysis consisted of the new $L=12,18,26$ data and a new sample for
$L=8$ calculated in the range $4/g^2=2.296-2.302$; the $L=36$ data were omitted
here, because close to the critical point still more statistics would have
been needed. As expected the better data produced a narrower $\chi^2-$parabola,
the minimum was shifted to a slightly smaller $4/g^2-$value, yet the 
1$\%-$difference of the exponent ratios remained. The 
apparent problem with the universality prediction disappeared however, 
when we included a correction-to-scaling
term like in eq. \ref{chi} into our fit, as it was done already in the case of 
the observable $g_r$. In fig. \ref{fig:crip} we compare the different minimal
$\chi^2-$curves for the observable $\MO$. In the $\omega$ dependent fits
and in the subsequent scaling analysis we
used as input the same set of critical exponents as \cite{Case97}. 
The corresponding minimal $\chi^2-$parabolae are even narrower than in the 
leading term fits, the result for 
$4/g^2_c$ is essentially independent of $\omega$ for $\omega=1.1-1.3$~ and  
is equal to the critical coupling found already in ref. \cite{Eng96}  
          \begin{equation}
          4/g_c^2 = 2.29895(10)~.
          \label{critc}
          \end{equation}
Using the observables $\chi_v$ and $g_r$ leads to similar, consistent results, 
preferring $\omega\approx 1.2$. 
          Here one should note that we describe the whole correction-to-scaling
          contributions with a single term. Consequently, the value of $\omega$
          is somewhat higher as expected from the relation 
          $\theta=\omega\nu=0.51(3)$ \cite{Case97}.


\subsection{The scaling functions}

\begin{figure}[htb]
\begin{center}
   \epsfig{bbllx=85,bblly=250,bburx=495,bbury=560,
       file=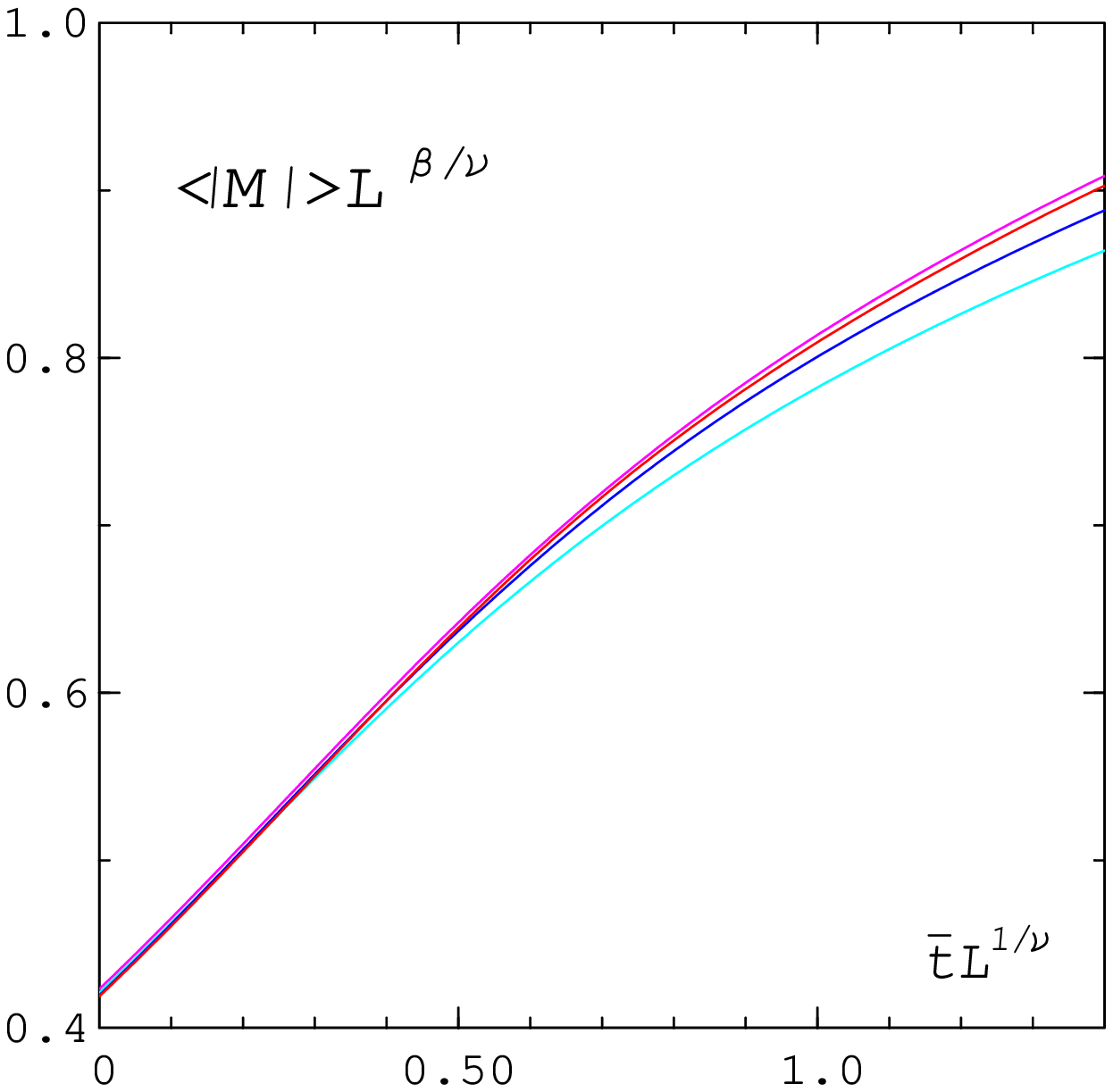, width=110mm,height=80mm}
\end{center}

\caption{The scaling function $\MO L^{\beta/\nu}$ vs. $x=\bar t L^{1/\nu}$ 
for $L=12$ (cyan), 18 (blue), 26 (red) and 36 (magenta).}

\label{fig:QM}
\end{figure}
\begin{figure}[htb]
\vskip 1.0truecm
\begin{center}
   \epsfig{bbllx=40,bblly=300,bburx=550,bbury=520,
       file=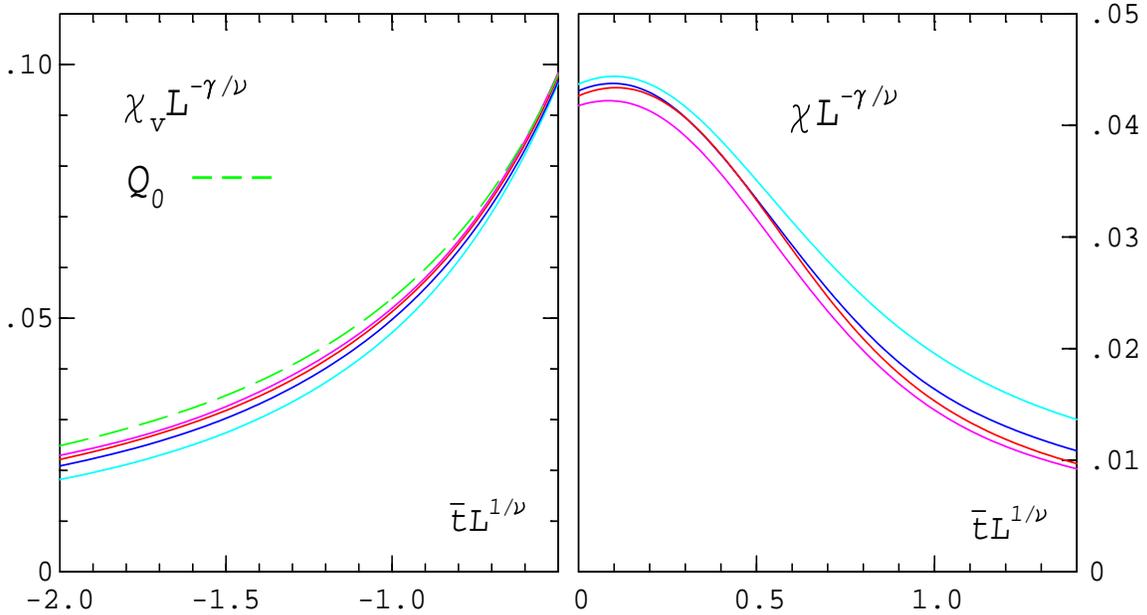, width=152mm,height=72mm}
\end{center}

\caption{The scaling functions $\chi_v L^{-\gamma/\nu}$ (left) and 
$\chi L^{-\gamma/\nu}$ (right) vs. $x=\bar t L^{1/\nu}$ 
for $L=12$ (cyan), 18 (blue), 26 (red) and 36 (magenta). The green curve shows
the estimated $Q_0(\chi_v)$ for $\omega=1.2$ .}

\label{fig:QCHI}
\end{figure}

 In Figs. \ref{fig:QM} and \ref{fig:QCHI}  we show the scaling functions 
$Q=L^{-\rho/\nu}O$ as a function of $x=\bar tL^{1/\nu}$ for $O=\MO,\chi_v$ 
and $\chi$. 
The remnant dependence of the scaling functions on the characteristic
length scale $L$ is due to corrections to scaling. A consistent
succession of curves at fixed $x$ for different $L$ emerged only after using
very high statistics and many couplings for the reweighting. 
To estimate $Q_0(x)$ we perform linear fits in
$y=L^{-\omega}$ of $Q(x,y)$ at fixed $x$. We find
a remarkable difference in the correction-to-scaling behaviours
in the two phases $\bar t<0$ and $\bar t>0$. In the symmetric 
phase, here for $\chi_v$, the correction-to-scaling contribution
is indeed linear in $y$. The best value of $\omega$ is again about 1.2.  
 In the broken 
phase the correction is certainly not linear in $y$ for small $L$, both in $Q_M$
and $Q_{\chi}$. Therefore we have estimated $Q_0$ here from the two largest 
lattices with $\omega$ in the range 1.1-1.3.
As can be seen from Fig. \ref{fig:QCHI} the signs
of the correction-to-scaling contributions are different for the
susceptibility in the two phases. The universal ratio for the 
correction-to-scaling amplitudes 
is therefore negative. From a high temperature expansion
Butera and Comi \cite{Bute98} predict that the correction amplitude of the
$N-$vector spin model $a_{1\chi}(N)$ is negative for $N\aleq 2$. Our finding of 
a negative correction-to-scaling contribution in the symmetric phase is in 
accord with this statement for $N=1$.

\noindent
Kiskis \cite{Joe92} considered another interesting scaling function. It is
defined by

\begin{equation}
h = {\OPQ \over \MO^2} -1.
\end{equation}

\begin{figure}[htb]
\begin{center}
   \epsfig{bbllx=85,bblly=250,bburx=495,bbury=560,
       file=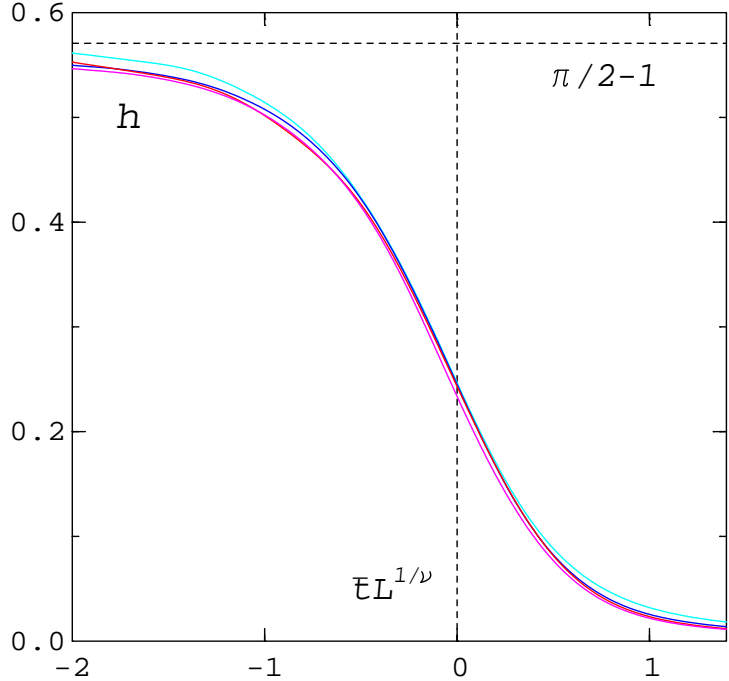, width=110mm,height=80mm}
\end{center}

\caption{The scaling function $h$ vs. $x=\bar t L^{1/\nu}$ 
for $L=12$ (cyan), 18 (blue), 26 (red) and 36 (magenta).}

\label{fig:hkis}
\end{figure}

\noindent
At the critical point $h$ is universal; there we find the value $h=0.240(5)$
from the $L=12,18,26$ lattices. In the
strong coupling limit $4/g^2 \rightarrow 0$ the function $h$ converges to
$\pi/2-1$, because then the distribution of the magnetization is Gaussian
\cite{Joe98}. This prediction can be checked in Fig. \ref{fig:hkis}. 
At a fixed negative $x-$value the smallest lattice is at the lowest 
$4/g^2-$value. Correspondingly the result on the lattice with $L=12$ 
reaches the predicted value earlier.  

\vskip 1.0truecm


\subsection{The critical amplitudes}

\begin{figure}[htb]
\begin{center}
   \epsfig{bbllx=85,bblly=250,bburx=495,bbury=560,
       file=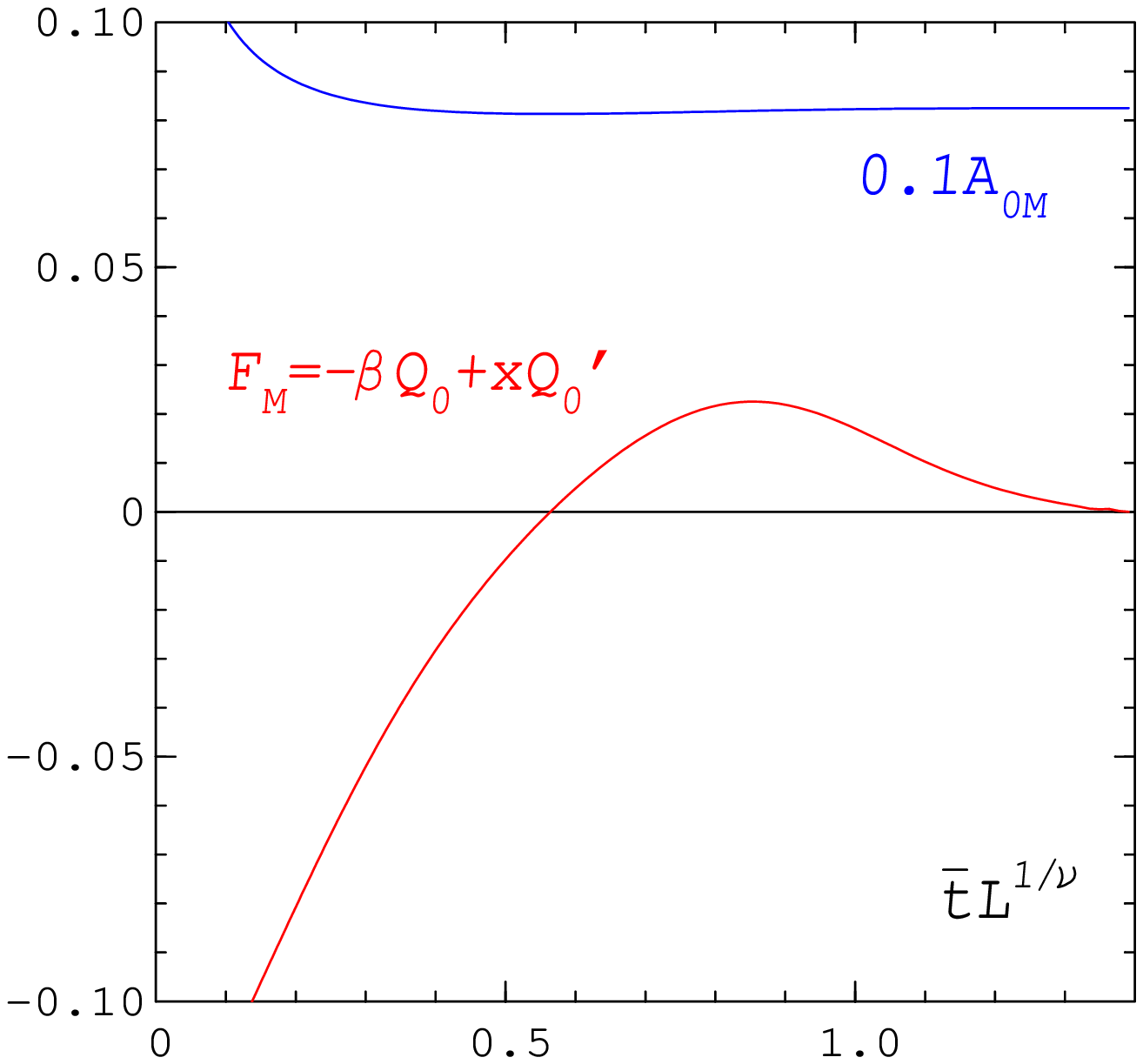, width=110mm,height=80mm}
\end{center}

\caption{The control function $F_M$ (red) and the amplitude function
$0.1A_{0M}$ (blue).}

\label{fig:FM}
\end{figure}
\begin{figure}[htb]
\vskip 1.0truecm
\begin{center}
   \epsfig{bbllx=85,bblly=250,bburx=495,bbury=560,
       file=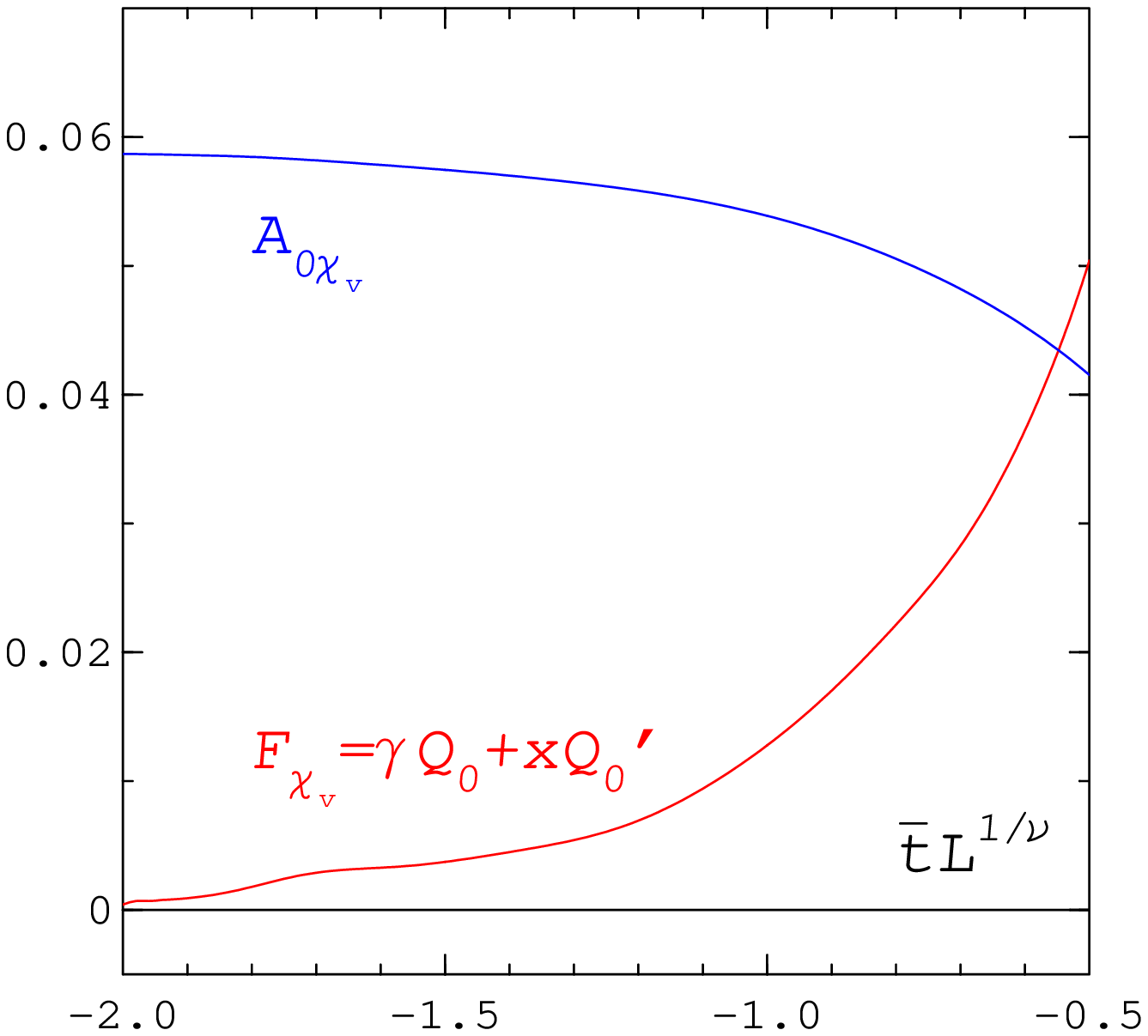, width=110mm,height=80mm}
\end{center}

\caption{The control function $F_{\chi_v}$ (red) and the amplitude function
$A_{0\chi_v}$ (blue).}

\label{fig:FCV}
\end{figure}
\begin{figure}[htb]
\begin{center}
   \epsfig{bbllx=85,bblly=250,bburx=495,bbury=560,
       file=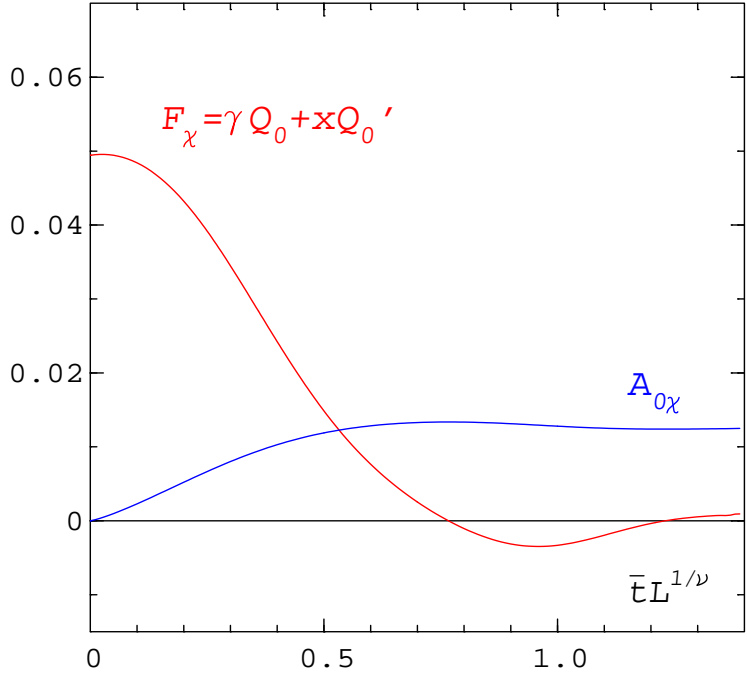, width=110mm,height=80mm}
\end{center}

\caption{The control function $F_{\chi}$ (red) and the amplitude function
$A_{0\chi}$ (blue).}

\label{fig:FC}
\end{figure}

In Figs. \ref{fig:FM}-\ref{fig:FC} we show the functions $F_O$, eq. \ref{Control},
which are obtained from the scaling functions $Q_0$ for $\MO,\chi_v$ and
$\chi$, respectively. In determining $Q_0$ the Jackknife errors of the
reweighted onservables were taken into account, the critical exponents 
from

\noindent the set \ref{crex} were used as input and $\omega$ was varied 
in the range
1.1-1.3 . The curves plotted in Figs. \ref{fig:FM}-\ref{fig:FC} correspond
to $\omega=1.2$ .We observe again a different behaviour below and above the
critical point. Whereas in the broken phase ($\bar t > 0$) both the control 
functions $F_M$ and $F_{\chi}$ have a single zero at small $x$ and become
essentially zero already around $x=1.4$, the approach to asymptopia in the 
symmetric phase (for $\chi_v$) is much slower. There the asymptotic region 
is reached only at $x\aleq -2$. The different behaviours are reflected as well
in the amplitude functions $A_0$ which are also shown in the figures. We may
now obtain the critical amplitudes from the amplitude functions in the
asymptotic domain where $F_O$ is compatible with zero. We find 

          \begin{eqnarray*}
          ~~~B~  & = & 0.825(1)~, \nonumber\\
          ~~~C_+ & = & 0.0587(8)~, \nonumber\\
          ~~~C_- & = & 0.01243(12)~.\nonumber\\
          \end{eqnarray*} 
\noindent
This amounts to a universal ratio for the susceptibility of
          \begin{equation}
           C_+/C_-  = 4.72(11)~.
          \end{equation} 
The errors in the amplitudes come from different sources. Apart from the 
errors in $A_0$ due to errors from the data, the main error comes from variations
in $\omega$ and errors from the point of onset of the asymptotic region. That
leads to a bigger error for $C_+$ than for the other quantities.
Our result for $C_+/C_-$ agrees nicely with the $3d$ Ising model value 4.75(3)
of \cite{Case97} and the latest field theoretic value 4.79(10) of \cite{GZJ98}.

 From our envelope formula \ref{enve} we can even derive an estimate for the 
next-to-leading amplitude 
          \begin{equation}
          a_1 = |x|^{-\omega\nu}Q_1(x)/Q_0(x)~.
          \end{equation} 
\noindent Though variations of $\omega$ influence strongly these 
correction-to-scaling amplitudes, their ratio is less affected. 
For the susceptibility we obtain the amplitude ratio $~a_{1+}/a_{1-}=-0.37(2)~$. 
As discussed already
this ratio is negative. The overall size of the ratio is however of comparable 
magnitude to other estimates \cite{ZJ}.


\section{Summary and conclusion}
 

 We have shown that it is possible to determine
 critical point amplitudes in $SU(2)$ from Monte Carlo simulations
 in finite, not extremely large volumes. 

  \noindent Very accurate data are however required for the necessary 
                 estimate of correction-to-scaling contributions to the
                 scaling functions and the control of their approach to
                 asymptopia.
    
  \noindent In the symmetric phase and in the broken phase we find 
  different correction-to-scaling dependencies of the scaling functions.

  \noindent Our result for $C_+/C_-$ is in excellent agreement with the $3d$
  Ising model value from Monte Carlo simulations and field 
  theory calculations of the $N=1-$vector model. 
  The agreement of this critical amplitude ratio for the (3+1) dimensional
  $SU(2)$ gauge theory and the $3d$ Ising model is a further strong support 
  of the  
  universality hypothesis of Svetitsky and Yaffe \cite{Svet82} beyond the
  level of critical exponents.

\newpage



\noindent{\large \bf Acknowledgements}


\noindent
We thank David Miller for a careful reading of the manuscript.



\newpage

\noindent{\large \bf Appendix}

\noindent In Tables 2-7 we present more details on our Monte Carlo simulations.
%
\begin{table}
\begin{center}
\begin{tabular}[t]{|c|c|c|rr|rr|c|}\hline
$4/g^2$  & $N_m$  &   $\langle |M| \rangle$   & $\chi_v$  & $\Delta \chi_v$ &  $\chi$ &$\Delta \chi$ & $g_r$\\ \hline\hline
2.28300  &  20000    &  0.0189    (02) &   25.7  &   0.6 &   9.05   &  0.21   &  -0.167   (45)\\ \hline
2.28400  &  30501    &  0.0198    (02) &   28.2  &   0.4 &   9.87   &  0.21   &  -0.240   (50)\\ \hline
2.28500  &  20000    &  0.0200    (02) &   29.0  &   0.6 &  10.43   &  0.26   &  -0.153   (34)\\ \hline
2.28600  &  30000    &  0.0210    (02) &   31.8  &   0.5 &  11.23   &  0.22   &  -0.176   (52)\\ \hline
2.28750  &  30000    &  0.0237    (02) &   39.6  &   0.6 &  13.51   &  0.24   &  -0.316   (48)\\ \hline
2.28850  &  30000    &  0.0248    (02) &   43.7  &   0.9 &  15.07   &  0.35   &  -0.287   (22)\\ \hline
2.29000  &  29436    &  0.0270    (05) &   51.4  &   1.6 &  17.48   &  0.54   &  -0.346   (44)\\ \hline
2.29120  &  32180    &  0.0294    (06) &   60.9  &   2.3 &  20.67   &  0.73   &  -0.368   (42)\\ \hline
2.29250  &  36751    &  0.0319    (04) &   70.3  &   1.4 &  22.85   &  0.42   &  -0.540   (51)\\ \hline
2.29500  &  26350    &  0.0409    (05) &  110.6  &   2.3 &  32.40   &  0.71   &  -0.818   (34)\\ \hline
2.29630  &  30000    &  0.0473    (08) &  143.1  &   3.8 &  38.72   &  0.82   &  -0.974   (36)\\ \hline
2.29750  &  40107    &  0.0553    (09) &  186.7  &   4.6 &  44.02   &  0.79   &  -1.191   (27)\\ \hline
2.29880  &  61070    &  0.0656    (11) &  247.5  &   6.7 &  46.64   &  0.54   &  -1.400   (21)\\ \hline
2.30000  &  57600    &  0.0738    (11) &  302.1  &   6.8 &  48.27   &  0.87   &  -1.522   (20)\\ \hline
2.30060  &  35000    &  0.0809    (11) &  351.5  &   7.5 &  45.99   &  0.96   &  -1.618   (15)\\ \hline
2.30130  &  36051    &  0.0881    (11) &  404.4  &   7.8 &  42.41   &  1.44   &  -1.699   (15)\\ \hline
2.30250  &  35622    &  0.0963    (10) &  472.4  &   7.7 &  39.53   &  1.94   &  -1.759   (13)\\ \hline
2.30320  &  41213    &  0.1023    (10) &  524.8  &   7.6 &  36.12   &  2.15   &  -1.806   (11)\\ \hline
2.30380  &  32000    &  0.1085    (10) &  577.9  &   8.2 &  28.91   &  2.22   &  -1.852   (11)\\ \hline
2.30440  &  41400    &  0.1113    (05) &  605.7  &   4.8 &  27.28   &  0.94   &  -1.861   (04)\\ \hline
2.30500  &  33600    &  0.1163    (05) &  655.7  &   5.1 &  24.46   &  1.39   &  -1.885   (05)\\ \hline
2.30600  &  20430    &  0.1224    (07) &  717.6  &   6.6 &  18.54   &  1.10   &  -1.914   (04)\\ \hline
2.30700  &  20000    &  0.1279    (06) &  778.8  &   6.0 &  15.57   &  0.95   &  -1.930   (04)\\ \hline
2.30800  &  20000    &  0.1334    (05) &  843.0  &   6.1 &  12.79   &  0.73   &  -1.945   (03)\\ \hline
2.30900  &  25513    &  0.1377    (02) &  896.5  &   2.6 &  11.55   &  0.31   &  -1.953   (01)\\ \hline
2.31000  &  20272    &  0.1422    (05) &  954.3  &   5.7 &  10.74   &  0.52   &  -1.959   (02)\\ \hline

\end{tabular}
\end{center}
\caption{The data from the $36^3\times4$ lattice. }
\label{tab:d36}
\end{table}
\newpage
\begin{table}
\begin{center}
\begin{tabular}{|c|c|c|rr|rr|c|}\hline
$4/g^2$  & $N_m$  &  $\langle |M| \rangle$  & $\chi_v$  & $\Delta \chi_v$ &  $\chi$ &$\Delta \chi$ & $g_r$\\ \hline\hline
2.27000  & 30000     & 0.0207    (2)  &  11.8   & 0.2   &  4.25  & 0.06    & -0.082    (38) \\ \hline
2.27250  & 30518     & 0.0221    (2)  &  13.4   & 0.2   &  4.77  & 0.07    & -0.121    (37) \\ \hline
2.27500  & 30043     & 0.0233    (2)  &  14.7   & 0.2   &  5.18  & 0.08    & -0.161    (44) \\ \hline
2.27750  & 20800     & 0.0254    (3)  &  17.5   & 0.4   &  6.14  & 0.15    & -0.215    (27) \\ \hline
2.28000  & 30000     & 0.0273    (2)  &  20.1   & 0.4   &  7.09  & 0.15    & -0.245    (14) \\ \hline
2.28250  & 20800     & 0.0298    (5)  &  23.9   & 0.7   &  8.23  & 0.24    & -0.270    (46) \\ \hline
2.28500  & 24204     & 0.0329    (4)  &  28.7   & 0.7   &  9.65  & 0.25    & -0.432    (38) \\ \hline
2.28750  & 20000     & 0.0369    (7)  &  35.6   & 1.3   & 11.74  & 0.35    & -0.472    (41) \\ \hline
2.29000  & 25000     & 0.0420    (5)  &  44.8   & 0.7   & 13.77  & 0.21    & -0.686    (37) \\ \hline
2.29200  & 30000     & 0.0470    (7)  &  55.6   & 1.4   & 16.83  & 0.27    & -0.744    (35) \\ \hline
2.29400  & 41525     & 0.0543    (8)  &  71.6   & 1.8   & 19.68  & 0.17    & -0.955    (28) \\ \hline
2.29600  & 60000     & 0.0626    (4)  &  91.6   & 1.0   & 22.76  & 0.29    & -1.134    (11) \\ \hline
2.29800  & 70000     & 0.0718    (4)  & 115.9   & 0.9   & 25.36  & 0.14    & -1.281    (08) \\ \hline
2.29900  & 40004     & 0.0782    (6)  & 132.6   & 1.6   & 25.19  & 0.27    & -1.405    (12) \\ \hline
2.30000  & 70000     & 0.0832    (4)  & 147.7   & 1.0   & 26.11  & 0.31    & -1.456    (09) \\ \hline
2.30120  & 59446     & 0.0922    (4)  & 174.0   & 1.6   & 24.64  & 0.28    & -1.582    (04) \\ \hline
2.30250  & 60761     & 0.0984    (7)  & 195.0   & 2.4   & 24.97  & 0.40    & -1.630    (09) \\ \hline
2.30370  & 60000     & 0.1069    (9)  & 224.3   & 2.6   & 23.57  & 0.71    & -1.703    (11) \\ \hline
2.30500  & 81780     & 0.1146    (6)  & 252.5   & 1.8   & 21.50  & 0.53    & -1.760    (06) \\ \hline
2.30620  & 60000     & 0.1218    (8)  & 279.2   & 2.8   & 18.58  & 0.71    & -1.808    (08) \\ \hline
2.30750  & 81059     & 0.1288    (5)  & 308.2   & 1.7   & 16.61  & 0.47    & -1.845    (04) \\ \hline
2.30870  & 40000     & 0.1352    (4)  & 334.7   & 1.4   & 13.39  & 0.54    & -1.876    (04) \\ \hline
2.31000  & 38400     & 0.1414    (5)  & 362.9   & 2.2   & 11.49  & 0.39    & -1.899    (03) \\ \hline
2.31120  & 60000     & 0.1461    (3)  & 385.7   & 1.1   & 10.36  & 0.47    & -1.913    (03) \\ \hline
2.31250  & 35000     & 0.1511    (3)  & 409.7   & 1.5   &  8.39  & 0.41    & -1.929    (02) \\ \hline
2.31380  & 20000     & 0.1553    (5)  & 431.9   & 2.6   &  8.05  & 0.39    & -1.936    (02) \\ \hline
2.31500  & 20000     & 0.1595    (6)  & 454.2   & 2.8   &  6.81  & 0.35    & -1.946    (03) \\ \hline
2.31750  & 20000     & 0.1672    (4)  & 497.3   & 2.0   &  5.84  & 0.24    & -1.957    (02) \\ \hline
2.32000  & 20000     & 0.1743    (2)  & 538.8   & 1.2   &  4.81  & 0.07    & -1.966    (00) \\ \hline 

\end{tabular}
\end{center}
\caption{The data from the $26^3\times4$ lattice. }
\label{tab:d26}
\end{table}

\newpage
\begin{table}
\begin{center}
\begin{tabular}{|c|c|c|r|r|c|}\hline
$4/g^2$  & $N_m$ & $\langle |M| \rangle$  & $\chi_v$\hspace{0.7cm}  &  $\chi$\hspace{0.71cm}  & $g_r$\\ \hline\hline
2.25000  & 20000     & 0.0253     (01) &   5.79 \hspace{0.3mm}    (08)  &  2.05     (04)    & -0.137    (23) \\ \hline 
2.25250  & 20000     & 0.0265     (01) &   6.33 \hspace{0.3mm}    (06)  &  2.24     (03)    & -0.208    (23) \\ \hline
2.25500  & 25000     & 0.0271     (02) &   6.64 \hspace{0.3mm}    (08)  &  2.36     (03)    & -0.146    (42) \\ \hline
2.25750  & 21500     & 0.0280     (02) &   7.10 \hspace{0.3mm}    (09)  &  2.53     (04)    & -0.127    (21) \\ \hline
2.26000  & 32816     & 0.0293     (02) &   7.78 \hspace{0.3mm}    (12)  &  2.77     (05)    & -0.127    (18) \\ \hline
2.26250  & 22497     & 0.0308     (02) &   8.52 \hspace{0.3mm}    (12)  &  2.99     (05)    & -0.200    (29) \\ \hline
2.26500  & 30000     & 0.0321     (02) &   9.21 \hspace{0.3mm}    (13)  &  3.21     (04)    & -0.212    (21) \\ \hline
2.26750  & 30857     & 0.0342     (02) &  10.40 \hspace{0.3mm}    (11)  &  3.58     (04)    & -0.302    (20) \\ \hline
2.27000  & 20000     & 0.0355     (03) &  11.22 \hspace{0.3mm}    (19)  &  3.86     (07)    & -0.286    (38) \\ \hline
2.27250  & 20200     & 0.0382     (04) &  12.84 \hspace{0.3mm}    (23)  &  4.35     (06)    & -0.353    (28) \\ \hline
2.27500  & 20000     & 0.0404     (04) &  14.38 \hspace{0.3mm}    (23)  &  4.87     (08)    & -0.382    (29) \\ \hline
2.27750  & 20000     & 0.0426     (05) &  15.78 \hspace{0.3mm}    (30)  &  5.20     (07)    & -0.489    (31) \\ \hline
2.28000  & 20000     & 0.0464     (03) &  18.60 \hspace{0.3mm}    (25)  &  6.03     (11)    & -0.539    (26) \\ \hline
2.28250  & 20000     & 0.0494     (03) &  20.95 \hspace{0.3mm}    (25)  &  6.72     (08)    & -0.598    (22) \\ \hline
2.28500  & 20000     & 0.0536     (05) &  24.35 \hspace{0.3mm}    (41)  &  7.59     (11)    & -0.673    (24) \\ \hline
2.28750  & 20490     & 0.0581     (05) &  28.06 \hspace{0.3mm}    (43)  &  8.40     (08)    & -0.771    (27) \\ \hline
2.29000  & 30000     & 0.0654     (10) &  34.61 \hspace{0.3mm}    (81)  &  9.64     (09)    & -0.934    (37) \\ \hline
2.29250  & 25000     & 0.0718     (06) &  40.58 \hspace{0.3mm}    (62)  & 10.49     (13)    & -1.067    (16) \\ \hline
2.29500  & 30075     & 0.0796     (05) &  48.40 \hspace{0.3mm}    (46)  & 11.49     (11)    & -1.182    (15) \\ \hline
2.29600  & 30000     & 0.0839     (06) &  52.92 \hspace{0.3mm}    (67)  & 11.85     (11)    & -1.261    (11) \\ \hline
2.29700  & 30000     & 0.0851     (09) &  54.38 \hspace{0.3mm}    (93)  & 12.18     (17)    & -1.260    (15) \\ \hline
2.29800  & 45000     & 0.0890     (09) &  58.48 \hspace{0.3mm}    (79)  & 12.32     (13)    & -1.315    (20) \\ \hline
2.29900  & 45000     & 0.0946     (06) &  64.72 \hspace{0.3mm}    (68)  & 12.59     (09)    & -1.387    (08) \\ \hline
2.30000  & 45000     & 0.0990     (06) &  69.71 \hspace{0.3mm}    (61)  & 12.61     (14)    & -1.441    (11) \\ \hline
2.30200  & 50151     & 0.1064     (10) &  78.73     (115) & 12.68     (24)    &
-1.518    (14) \\ \hline
\end{tabular}
\end{center}
\caption{List(a) of data from the $18^3\times4$ lattice. }
\label{tab:d18a}
\end{table}
\newpage
\begin{table}
\begin{center}
\begin{tabular}{|c|c|c|r|r|c|}\hline
$4/g^2$  & $N_m$ & $\langle |M| \rangle$  & $\chi_v$\hspace{0.7cm}  &  $\chi$\hspace{0.7cm}  & $g_r$\\ \hline\hline
2.30350  & 50856     & 0.1113     (08) &  84.92 \hspace{0.3mm}    (85)  & 12.74     (27)    & -1.560    (14) \\ \hline
2.30500  & 49940     & 0.1208     (09) &  96.95     (106) & 11.85     (19)    & -1.643    (09) \\ \hline
2.30600  & 39257     & 0.1232     (06) & 100.36 \hspace{0.3mm}    (73)  & 11.87     (17)    & -1.660    (07) \\ \hline
2.30700  & 20157     & 0.1269     (08) & 105.25     (107) & 11.40     (30)    & -1.686    (09) \\ \hline
2.30850  & 40574     & 0.1336     (08) & 115.04     (101) & 10.97     (34)    & -1.728    (08) \\ \hline
2.31000  & 40000     & 0.1395     (09) & 123.64     (117) & 10.15     (28)    & -1.765    (07) \\ \hline
2.31100  & 20000     & 0.1434     (12) & 129.55     (165) & 9.70     (35)    & -1.785    (08) \\ \hline
2.31300  & 40000     & 0.1506     (05) & 140.89 \hspace{0.3mm}    (72)  & 8.70     (23)    & -1.820    (03) \\ \hline
2.31500  & 30000     & 0.1586     (04) & 154.14 \hspace{0.3mm}    (52)  & 7.39     (24)    & -1.854    (03) \\ \hline
2.31650  & 25000     & 0.1617     (06) & 159.70 \hspace{0.3mm}    (87)  & 7.21     (29)    & -1.863    (04) \\ \hline
2.31800  & 20115     & 0.1668     (05) & 168.54 \hspace{0.3mm}    (81)  & 6.32     (22)    & -1.882    (02) \\ \hline
2.32000  & 30000     & 0.1728     (05) & 179.55 \hspace{0.3mm}    (91)  & 5.37     (19)    & -1.902    (03) \\ \hline
2.32200  & 22721     & 0.1780     (05) & 189.71 \hspace{0.3mm}    (79)  & 4.98     (29)    & -1.914    (03) \\ \hline
2.32500  & 20000     & 0.1851     (03) & 203.80 \hspace{0.3mm}    (53)  & 4.03     (11)    & -1.931    (01) \\ \hline
2.32750  & 20000     & 0.1900     (02) & 214.13 \hspace{0.3mm}    (38)  & 3.67     (09)    & -1.939    (01) \\ \hline
2.33000  & 21071     & 0.1949     (02) & 225.07 \hspace{0.3mm}    (38)  & 3.53     (13)    & -1.945    (01) \\ \hline
2.33250  & 20000     & 0.1992     (02) & 234.60 \hspace{0.3mm}    (48)  & 3.09     (08)    & -1.952    (01) \\ \hline
2.33500  & 20015     & 0.2042     (02) & 245.93 \hspace{0.3mm}    (35)  & 2.85     (03)    & -1.957    (00) \\ \hline
2.33750  & 20000     & 0.2080     (02) & 254.92 \hspace{0.3mm}    (49)  & 2.57     (05)    & -1.962    (01) \\ \hline
2.34000  & 20000     & 0.2118     (01) & 264.19 \hspace{0.3mm}    (31)  & 2.51     (05)    & -1.965    (00) \\ \hline
2.34250  & 20000     & 0.2154     (03) & 273.05 \hspace{0.3mm}    (62)  & 2.40     (04)    & -1.967    (01) \\ \hline
2.34500  & 20026     & 0.2191     (02) & 282.24 \hspace{0.3mm}    (48)  & 2.18     (02)    & -1.970    (00) \\ \hline
2.34750  & 20000     & 0.2224     (02) & 290.62 \hspace{0.3mm}    (43)  & 2.15     (01)    & -1.972    (00) \\ \hline
2.35000  & 20086     & 0.2257     (02) & 299.15 \hspace{0.3mm}    (40)  & 2.00     (04)    & -1.974    (00) \\ \hline
\end{tabular}
\end{center}
\caption{List(b) of data from the $18^3\times4$ lattice. }
\label{tab:d18b}
\end{table}

\renewcommand{\baselinestretch}{0.85}

\newpage
\begin{table}
\begin{center}
\begin{tabular}{|c|c|c|r|r|c|}\hline
$4/g^2$  & $N_m$  & $\langle |M| \rangle$   & $\chi_v$\hspace{0.7cm}  & $\chi$\hspace{0.6cm}  & $g_r$\\ \hline\hline
  2.20500  &   20000   &   0.0286   (1) &   2.22     (01) &  0.80  (1) &   -0.058   (50)\\ \hline 
  2.20750  &   20000   &   0.0294   (2) &   2.34     (03) &  0.85  (1) &   -0.012   (34)\\ \hline
  2.21000  &   20460   &   0.0300   (1) &   2.41     (02) &  0.85  (1) &   -0.123   (43)\\ \hline
  2.21250  &   20000   &   0.0304   (2) &   2.51     (03) &  0.91  (1) &   -0.026   (39)\\ \hline
  2.21500  &   20000   &   0.0311   (2) &   2.61     (03) &  0.94  (1) &   -0.109   (39)\\ \hline
  2.21750  &   20000   &   0.0319   (1) &   2.75     (02) &  0.99  (1) &   -0.081   (25)\\ \hline
  2.22000  &   20000   &   0.0326   (2) &   2.87     (04) &  1.03  (1) &   -0.069   (57)\\ \hline
  2.22250  &   20512   &   0.0335   (2) &   3.00     (03) &  1.06  (1) &   -0.166   (29)\\ \hline
  2.22500  &   20000   &   0.0339   (2) &   3.09     (03) &  1.11  (1) &   -0.085   (31)\\ \hline
  2.22750  &   20000   &   0.0350   (1) &   3.28     (02) &  1.17  (1) &   -0.114   (22)\\ \hline
  2.23000  &   20000   &   0.0362   (1) &   3.51     (02) &  1.25  (1) &   -0.156   (32)\\ \hline
  2.23250  &   20498   &   0.0370   (2) &   3.66     (04) &  1.29  (1) &   -0.172   (14)\\ \hline
  2.23500  &   20014   &   0.0382   (3) &   3.88     (05) &  1.36  (2) &   -0.224   (28)\\ \hline
  2.23750  &   20000   &   0.0392   (3) &   4.08     (05) &  1.43  (1) &   -0.207   (36)\\ \hline
  2.24000  &   20000   &   0.0406   (2) &   4.36     (04) &  1.52  (1) &   -0.250   (35)\\ \hline
  2.24250  &   20000   &   0.0422   (2) &   4.71     (05) &  1.64  (2) &   -0.261   (19)\\ \hline
  2.24500  &   20000   &   0.0431   (2) &   4.92     (05) &  1.71  (3) &   -0.284   (31)\\ \hline
  2.24750  &   20000   &   0.0443   (3) &   5.20     (06) &  1.81  (2) &   -0.261   (23)\\ \hline
  2.25000  &   20000   &   0.0459   (3) &   5.54     (06) &  1.90  (2) &   -0.328   (06)\\ \hline
  2.25250  &   20000   &   0.0470   (4) &   5.81     (08) &  1.99  (3) &   -0.347   (32)\\ \hline
  2.25500  &   20000   &   0.0488   (2) &   6.27     (06) &  2.16  (3) &   -0.329   (14)\\ \hline
  2.25750  &   20000   &   0.0513   (3) &   6.83     (05) &  2.28  (1) &   -0.449   (23)\\ \hline
  2.26000  &   20974   &   0.0533   (4) &   7.34     (07) &  2.43  (2) &   -0.453   (26)\\ \hline
  2.26250  &   20000   &   0.0556   (4) &   7.95     (11) &  2.61  (3) &   -0.498   (14)\\ \hline
  2.26500  &   20000   &   0.0576   (5) &   8.52     (12) &  2.79  (3) &   -0.529   (17)\\ \hline
  2.26750  &   26000   &   0.0607   (4) &   9.29     (10) &  2.92  (3) &   -0.603   (11)\\ \hline
  2.27000  &   20000   &   0.0640   (4) &   10.29    (12) &  3.21  (3) &   -0.660   (18)\\ \hline
  2.27250  &   20000   &   0.0669   (6) &   11.16    (16) &  3.43  (3) &   -0.697   (21)\\ \hline
  2.27500  &   20000   &   0.0690   (3) &   11.83    (11) &  3.61  (4) &   -0.724   (17)\\ \hline
  2.27750  &   20000   &   0.0730   (7) &   13.07    (21) &  3.87  (4) &   -0.803   (21)\\ \hline
  2.28000  &   20056   &   0.0764   (3) &   14.22    (11) &  4.14  (4) &   -0.839   (09)\\ \hline
  2.28250  &   20000   &   0.0814   (8) &   15.85    (27) &  4.40  (4) &   -0.931   (16)\\ \hline
  2.28500  &   20415   &   0.0858   (6) &   17.40    (20) &  4.67  (4) &   -1.009   (18)\\ \hline
  2.28750  &   20000   &   0.0904   (7) &   18.97    (21) &  4.86  (4) &   -1.079   (17)\\ \hline
  2.29000  &   40000   &   0.0952   (3) &   20.78    (09) &  5.13  (4) &   -1.129   (09)\\ \hline
  2.29250  &   40000   &   0.1014   (8) &   23.17    (29) &  5.39  (4) &   -1.207   (13)\\ \hline
  2.29500  &   40000   &   0.1066   (9) &   25.12    (30) &  5.50  (4) &   -1.277   (15)\\ \hline
\end{tabular}
\end{center}
\caption{List(a) of data from the $12^3\times4$ lattice. }
\label{tab:d12a}
\end{table}
\newpage
\begin{table}
\begin{center}
\begin{tabular}{|c|c|c|r|r|c|}\hline
$4/g^2$  & $N_m$  & $\langle |M|\rangle$  & $\chi_v$\hspace{0.7cm}  &  $\chi$\hspace{0.7cm}  & $g_r$\\ \hline\hline
  2.29700  &   40000   &   0.1105  (6) &   26.80   (22)  &  5.70   (04)  &   -1.308   (09)\\ \hline
  2.29800  &   40000   &   0.1129  (8) &   27.66   (27)  &  5.62   (04)  &   -1.344   (11)\\ \hline
  2.29900  &   50000   &   0.1165  (5) &   29.19   (17)  &  5.74   (03)  &   -1.375   (07)\\ \hline
  2.30000  &   30000   &   0.1193  (6) &   30.29   (20)  &  5.71   (05)  &   -1.407   (08)\\ \hline
  2.30200  &   30107   &   0.1234  (8) &   32.16   (35)  &  5.84   (05)  &   -1.439   (10)\\ \hline
  2.30500  &   30000   &   0.1326  (7) &   36.17   (29)  &  5.78   (05)  &   -1.522   (08)\\ \hline
  2.30700  &   50000   &   0.1362  (7) &   37.90   (31)  &  5.86   (05)  &   -1.540   (07)\\ \hline
  2.31000  &   30000   &   0.1447  (9) &   41.85   (37)  &  5.66   (09)  &   -1.604   (09)\\ \hline
  2.31300  &   20000   &   0.1546  (8) &   46.66   (37)  &  5.35   (09)  &   -1.668   (07)\\ \hline
  2.31500  &   20000   &   0.1592  (8) &   48.96   (35)  &  5.15   (12)  &   -1.694   (07)\\ \hline
  2.31700  &   20000   &   0.1636  (12) &  51.26   (56)  &  5.01   (15)  &   -1.717   (09)\\ \hline
  2.31900  &   20000   &   0.1687  (5) &   53.96    (24) &  4.79   (10)  &   -1.742   (04)\\ \hline
  2.32000  &   20000   &   0.1707  (8) &   55.11    (40) &  4.75   (08)  &   -1.750   (04)\\ \hline
  2.32250  &   20000   &   0.1754  (8) &   57.66    (40) &  4.53   (06)  &   -1.772   (04)\\ \hline
  2.32500  &   39432   &   0.1814  (5) &   61.15    (26) &  4.27   (06)  &   -1.795   (03)\\ \hline
  2.32750  &   40000   &   0.1879  (5) &   64.94    (25) &  3.92   (08)  &   -1.821   (03)\\ \hline
  2.33000  &   40000   &   0.1917  (3) &   67.30    (18) &  3.80   (04)  &   -1.832   (01)\\ \hline
  2.33250  &   20000   &   0.1976  (4) &   70.84    (24) &  3.34   (04)  &   -1.854   (02)\\ \hline
  2.33500  &   40000   &   0.2019  (5) &   73.54    (30) &  3.13   (06)  &   -1.866   (03)\\ \hline
  2.33750  &   39618   &   0.2063  (4) &   76.59    (25) &  3.02   (06)  &   -1.876   (02)\\ \hline
  2.34000  &   20000   &   0.2106  (5) &   79.38    (35) &  2.74   (06)  &   -1.888   (02)\\ \hline
  2.34250  &   20000   &   0.2139  (3) &   81.72    (20) &  2.63   (07)  &   -1.896   (02)\\ \hline
  2.34500  &   18371   &   0.2180  (2) &   84.60    (14) &  2.48   (07)  &   -1.904   (01)\\ \hline
  2.34750  &   20000   &   0.2210  (3) &   86.76    (20) &  2.35   (06)  &   -1.908   (02)\\ \hline
  2.35000  &   20000   &   0.2249  (2) &   89.62    (17) &  2.19   (05)  &   -1.916   (01)\\ \hline
  2.35250  &   20000   &   0.2283  (4) &   92.19    (30) &  2.13   (06)  &   -1.921   (02)\\ \hline
  2.35500  &   20000   &   0.2305  (3) &   93.86    (20) &  2.07   (06)  &   -1.924   (01)\\ \hline
  2.35750  &   20000   &   0.2342  (3) &   96.65    (20) &  1.87   (03)  &   -1.931   (01)\\ \hline
  2.36000  &   20000   &   0.2365  (4) &   98.47    (24) &  1.85   (06)  &   -1.933   (02)\\ \hline
  2.36250  &   20000   &   0.2395  (3) &  100.89    (24) &  1.79   (07)  &   -1.937   (02)\\ \hline
  2.36500  &   20000   &   0.2414  (3) &  102.50    (25) &  1.77   (06)  &   -1.939   (01)\\ \hline
  2.36750  &   20222   &   0.2441  (2) &  104.61    (19) &  1.67   (04)  &   -1.942   (01)\\ \hline
  2.37000  &   20000   &   0.2467  (3) &  106.80    (24) &  1.63   (04)  &   -1.945   (01)\\ \hline
  2.37250  &   20000   &   0.2493  (3) &  108.95    (21) &  1.53   (03)  &   -1.948   (01)\\ \hline
  2.37500  &   20000   &   0.2515  (3) &  110.75    (23) &  1.47   (02)  &   -1.950   (01)\\ \hline
  2.37750  &   20000   &   0.2543  (2) &  113.19    (21) &  1.44   (02)  &   -1.953   (00)\\ \hline
  2.38000  &   20000   &   0.2563  (3) &  114.96    (24) &  1.41   (02)  &   -1.954   (01)\\ \hline
\end{tabular}

\end{center}
\caption{List(b) of data from the $12^3\times4$ lattice. }
\label{tab:d12b}
\end{table}





\begin{thebibliography}{9}
\bibitem{AHP91} V.~Privman, P.C.~Hohenberg, A.~Aharony, {\it Universal~
Critical~Point~Amplitude \break Relations}, in "Phase Transitions and Critical
Phenomena", vol. 14, C. Domb and J.L. Lebowitz eds. (Academic Press 1991).  
\bibitem{Case97} M.~Caselle and M.~Hasenbusch, J. Phys. A30 (1997) 4963;
                 \NP B(Proc.Suppl.)63 (1998) 613.
\bibitem{BLZ74} E.~Brezin, J.-C.~LeGuillou and J.~Zinn-Justin, \PL 47A (1974) 285.
\bibitem{Ahar76} A.~Aharony and P.C.~Hohenberg, \PR B13 (1976) 3081.
\bibitem{Albr85} P.C.~Albright and J.F.~Nicoll, \PR B31 (1985) 4576.
\bibitem{Berv86} C.~Bervillier, \PR B34 (1986) 8141.
\bibitem{Bagn87} C.~Bagnuls, C.~Bervillier, D.I.~Meiron and B.G.~Nickel, \PR
B35 (1987)3585.
\bibitem{GZJ97} R.~Guida and J.~Zinn-Justin, \NP B489[FS] (1997) 626.
\bibitem{GZJ98} R.~Guida and J.~Zinn-Justin,{\it Critical exponents of the  
                N-vector model}, Preprint SPhT-t97/040, cond-mat/9803240 v2.
\bibitem{Bute98} P.~Butera and M.~Comi, hep-lat/9805025.  
\bibitem{Joe92}  J.~Kiskis, \PR D45 (1992) 4640.
\bibitem{Eng96}  J.~Engels, S.~Mashkevich, T.~Scheideler and G.~Zinovjev,
 \PL B365 (1996) 219.
\bibitem{Barb83} M.N.~Barber, in "Phase Transitions and Critical
Phenomena", vol. 8, C. Domb and J.L. Lebowitz eds. (Academic Press 1983).  
\bibitem{Tala96} A.L.~Talapov and H.W.J.~Bl\"ote, J. Phys. A29 (1996) 5727.
\bibitem{Eng92}  J.~Engels, J.~Fingberg and D.E.~Miller, \NP B387 (1992) 501.
\bibitem{DSM} M.~Falconi, E.~Marinari, M.L.~Paciello, 
G.~Parisi and B.~Taglienti, \PL 108B (1982) 331;
E.~Marinari, \NP B235 (1984) 123;\\
G.~Bhanot, S.~Black, P.~Carter and R.~Salvador, \PL 183B (1986) 331;\\
A.M. Ferrenberg and R.H. Swendsen,  Phys. Rev. Lett. 61
(1988) 2635; 63 (1989) 1195.
\bibitem{Joe98}  J.~Kiskis, private communication.
\bibitem{ZJ} J.~Zinn-Justin,{\it~Quantum field theory and critical
                phenomena} (1993) Clarendon Press, Oxford.
\bibitem{Svet82} B.~Svetitsky and G.~Yaffe, \NP B210{FS6} (1982) 423.
\end{thebibliography}
\end{document}